\documentclass[11pt,preprint2]{aastex}

\usepackage{color}
\usepackage{paralist}
\usepackage{longtable}
\usepackage{multirow}
\usepackage{booktabs}

\slugcomment{}

\shorttitle{Impulsivity Parameter for Solar Flares}
\shortauthors{Fajardo-Mendieta, Mart\'inez Oliveros, Alvarado-G\'omez and Calvo-Mozo}

\newcommand{\ignore}[1]{}

\begin{document}


\title{\MakeUppercase{Impulsivity Parameter for Solar Flares}}

\author{\textsc {W.G. Fajardo-Mendieta$^{1,2}$, J.C. Mart\'{i}nez-Oliveros$^{3}$, J.D. Alvarado-G\'{o}mez$^{1,4,5}$, B. Calvo-Mozo$^{1}$}}
\affil{$^1$Observatorio Astron\'omico Nacional, Universidad Nacional de Colombia, Bogot\'a, Colombia; wgfajardom@unal.edu.co, bcalvom@unal.edu.co\\
$^2$Departamento de F\'isica, Universidad Nacional de Colombia, Bogot\'a, Colombia\\
$^3$Space Sciences Laboratory, UC Berkeley, Berkeley, CA 94720, USA; oliveros@ssl.berkeley.edu\\
$^4$European Southern Observatory, Karl-Schwarzschild-Str. 2, D-85748 Garching bei M\"unchen, Germany; jalvarad@eso.org\\
$^5$Universit\"ats-Sternwarte M\"unchen, Ludwig-Maximilians-Universit\"at, Scheinerstr.~1, D-81679 M\"unchen, Germany}



\begin{abstract}
Three phases are typically observed during solar flares: the preflare, impulsive, and decay phases. During the impulsive phase, it is believed 
that the electrons and other particles are accelerated after the stored energy in the magnetic field is released by reconnection. 
The impulsivity of a solar flare is a quantifiable property that shows how quickly this initial energy release occurs. 
It is measured via the impulsivity parameter, which we define as the inverse of the overall duration of the impulsive phase. 
We take the latter as the raw width of the most prominent nonthermal emission of the flare.
We computed this observable over a work sample of 48 M-class events that occurred during the current Solar Cycle 24 by 
using three different methods. The first method takes into account all of the nonthermal flare emission and gives very accurate results, 
while the other two just cover fixed energy intervals (30-40 keV and 25-50 keV) and are useful for fast calculations.
We propose an alternative way to classify solar flares according to their impulsivity parameter values, defining three different types of impulsivity, 
namely, high, medium, and low. This system of classification is independent of the manner used to calculated the impulsivity parameter. 
Lastly, we show the relevance of this tool as a discriminator of different HXR generation processes.
\end{abstract}



\section{Introduction}


The sudden and localized releases of energy stored in the magnetic field and their subsequent effects on the solar atmosphere are known as solar flares. 
The current standard model for solar flares \citep{Hirayama1974,CargillPriest1983} considers magnetic reconnection as the physical process 
responsible for such energy release. The stored non-potential magnetic energy is transformed into kinetic and thermal energy, which is used to heat 
the surrounding plasma, and to accelerate particles adjacent to the reconnection region toward the outer and inner layers of the solar atmosphere. 
The former could be expelled out of the Sun (Solar Energetic Particle events - SEP, Coronal Mass Ejections - CME), while the latter form the 
primary beam of accelerated particles, which descends along magnetic field lines and deposits the majority of its energy in the denser layers, as 
the low chromosphere and high photosphere. Synchrotron radiation, bremsstrahlung, and white-light emissions are typical during this initial evolutionary 
stage \citep{Brown1971,Emslie1978}.

As the primary beam of accelerated particles is thermalized, the chromospheric plasma is rapidly heated to temperatures higher than the 
surrounding material. The plasma cannot dissipate the incoming energy, it therefore rises through the loop populating it with hot material. 
This causes increases in the average pressure and density in the loop. This process is known as {\it chromospheric evaporation} \citep{Antonucci1982}. 
Once the chromospheric material is evaporated into the lower corona, the plasma inside the loop relaxes and its temperature decreases, shifting its
dominant emission from soft X-rays (SXR) to H$\alpha$. Finally, the remaining flare energy is thermally and radiatively dissipated. At this 
last stage, the whole system becomes quieter than before, except for the upper corona, where shock waves proceed into interplanetary space 
creating radio bursts and accelerating other particles \citep{Benz2008}.


The temporal evolution of solar flares has been taxonomically organized into a set of four phases, namely, preflare, impulsive, flash, and 
decay \citep{Benz2002}. We are interested in the impulsive phase because, in this stage, most of the flare energy is deposited in the system \citep{Hudson2011}.
Thus, it can be the cornerstone for the understanding of the subsequent evolutionary phases of solar flares.


Historically, the study of the impulsive phase has been focused primarily in three topics:
the overall balance of flare energy \citep[e.g.][]{Antonucci1982,Fletcher2013}, 
the measurement of the plasma properties in flaring conditions, \citep[e.g.][]{Graham2011,Graham2013}, and 
the correlation with other types of solar activity \citep{Zhang2001}.
There is an additional area that also has a long tradition: the description of the impulsivity \citep[e.g.][]{Crannell1978, Abbett1999}.
However, this concept has so far only been discussed from a qualitative point of view.


The intuitive ideas about impulsivity come mostly from observational studies of hard X-rays (HXR). These kinds of emissions 
are believed to be generated in the impulsive phase during the thermalization of the primary beam \citep{Benz2008}.
\citet{Grigis2004} and \citet{Kiplinger1995} showed that the spectral index of the nonthermal X-ray spectrum of a solar flare can 
follow two different evolutionary empirical patterns known as {\it soft-hard-soft} and {\it soft-hard-harder}, respectively. The {\it soft-hard-soft} 
spectral shape is related to those flares whose HXR emissions are confined to a single pulse and occur in a time interval of the order 
of seconds, while the {\it soft-hard-harder} pattern corresponds to events whose HXR emissions are distributed in multiple peaks and/or have
a long duration, ranging from tens of seconds to even a couple of minutes. The first kind of flares are called {\it impulsive} and the second kind 
are called {\it gradual} or {\it Long Duration Events - (LDE)} \citep{Shibata1996,Dennis1989}.


Additionally, this flare classification is supported by one of the oldest observational correlations that exists between two different wavelenghts in 
solar physics: the Neupert effect \citep{Neupert1968,Hudson1991}. The Neupert effect says that the accumulated flux of a solar flare in HXR, i.e. 
integrated over some time interval, is equivalent, or at least proportional, to the flux of the same event in SXR at a particular time.
This can be also expressed as \citep{Veronig2005}:

\begin{equation}
\frac{d}{dt}F_{SXR}(t) \propto F_{HXR}(t)
\end{equation}

\citet{Dennis1993} computed in what proportion the Neupert effect holds over a sample of 92 flares, from which 66 were impulsive and 26 were LDE. They obtained
that impulsive flares fulfilled the Neupert effect in a larger proportion in comparison with LDE events (a ratio of 53/66 in contrast to 12/26).


These works show the current understanding of the impulsivity concept. 
Nevertheless, we wonder the following: could the impulsivity be something more than a qualitative or taxonomical feature of solar flares? 
This question highlights the aim of this article. Here, we consider the impulsivity as a quantitative characteristic, and we propose 
a method to measure it via a new observational parameter. This text is divided into four more sections. In the next section, we define the impulsivity 
parameter and discuss its physical meaning. Section~\ref{da} will be devoted to the description of the flare sample and a detailed explanation
of the impulsivity parameter calculation. In Section~\ref{re}, we show the main results of our research, e.g. the alternative system of classification 
for solar flares based on the impulsivity parameter. Finally, we discuss the restrictions and possible applications of this new tool.

\section{Definition of the Impulsivity Parameter}
\label{ip}


Nowadays, it is believed that the impulsive phase is characterized by the release of the primary electron beam and its corresponding process of thermalization along the loop \citep{Benz2008}. 
These processes occur over very short time intervals. Usually, microwaves and HXR emissions are observed during the impulsive phase of the flare \citep{Ning2007}. 
However, how can one define how impulsive a flare is? It goes beyond the intuitive and qualitative concept of impulsivity that we talked about in the 
Introduction. Here we define the impulsivity as a measurable quantity, which can be compared between different flares. This feature, of 
course, must be common to all flares.


We propose taking the flare impulsivity as inversely proportional to the duration of the impulsive phase.
Additionally, this measurable physical quantity is a key
feature of solar flares that may give us information about their later behavior. For example, two flares of the same GOES class could 
have completely different temporal evolutions, whether the duration of their impulsive phases differs by an order of magnitude. The 
energy input rates would probably have different values, and hence the effects over the surrounding neighborhoods would not be the same.

Although the temporal evolution of solar flares is well defined conceptually, the boundaries between one phase and another are very diffuse observationally. 
For this reason, we design a method to estimate the impulsivity based on the emission that occurs predominantly during the impulsive phase. We consider the 
duration of the impulsive phase as the duration of the most intense nonthermal HXR emission produced by the flare (see section~\ref{da}). 
The ratio between a normalization factor (defined at Section \ref{da}) and the impulsive phase duration is what we will call the impulsivity parameter.


We could choose other physical properties of the nonthermal emissions as candidates for the impulsivity parameter, e.g. 
the total fluence or just the rise time of the most 
prominent peak. However, they present some problems that make them less desirable. The total fluence of nonthermal emissions accounts for only a small 
fraction of the total flare energy, non-potential magnetic energy, and it varies significantly from one event to another \citep{Emslie2012}. Therefore, 
we cannot ensure that it is directly proportional to the impulsivity. On the other hand, the rise time gives us the duration of the most powerful injection, 
but we are missing the relaxation or decay time of such emission, which could be also interesting, as we will see in the next section.


In general, nonthermal HXR emission comes from the footpoints. 
This radiation is mostly produced via collisions between atmospheric plasma of high density and the primary electron beam, 
which began its travel from the acceleration region near the reconnection point. 
Here, for simplicity, we neglect the contributions from any other flare related HXR sources.\footnote{\noindent In 
some events we see a third HXR kernel of emission located above the loop \citep[e.g.][]{Masuda1994,Krucker2008,Glesener2013}.} 
Therefore, we can establish a direct relation of proportionality between the duration of the emission in the footpoints and the 
lifetime of the primary electron beam. Thus, by measuring the duration of the impulsive phase, we may approach another scientific question, one even 
more interesting: what is the lifetime of the primary electron beam? Of course, we must be aware that this relation holds only under the assumptions made 
above, i.e. the bulk of the HXR emission is released in a single pulse and footpoints are the main emitters.

\section{Data and Analysis}
\label{da}

The flare sample, over which the impulsivity parameter was computed, was chosen according to the following ad hoc criteria. 
The events took place from 2008 to 2013, with a GOES class between M2.0 and M9.9, and were observed by the {\it Reuven Ramaty High 
Energy Solar Spectroscopic Imager (RHESSI)} \citep{Lin2002}, during at least 90\% of the rise time of their corresponding GOES lightcurves.
The first two criteria were fullfilled by 131 flares; however, the latter condition greatly reduces the size of the sample, due to  
instrument eclipses, and the spacecraft passes through the South Atlantic Anomaly (SAA). Finally, only 48 events satisfied these conditions. 
They compose our work sample. 

We leave out flares of GOES class C because they have a high probability of not presenting a clear prominent peak in their nonthermal X-ray lightcurves and under 
this condition the impulsivity parameter would be extremely difficult to measure. This assumption is based on the fact that these events have very low photon 
fluxes at nonthermal energies. Additionally, we do not consider flares of the GOES class ranging between M1.0 and M1.9, 
because by including them the sample size would have increased by 270\%, making data analysis hard to handle. 
These events could be considered in a future systematic study following the same procedure as for the ones considered in this paper.

On the other hand, one highly energetic event detected by GOES could be generated by the combination of many smaller or weaker flares. In principle, we want to
analyze one strong nonthermal emission for each GOES event, but in these cases we do not know over which of those prominent peaks we should compute the 
impulsivity parameter. In order to avoid this degeneration problem, we did not include X class flares in this study.

The impulsivity parameter was calculated in all the flares of the work sample following the same methodology. We will use one event as  
an example in order to explain this methodology step by step, namely the  event (SOL2012-06-03T17:55-M3.3). The analysis time interval taken spans from 10 
minutes before the beginning of the flare to 10 minutes after the end of the flare, both times according to GOES, i.e. from 17:38:00 to 18:07:00.
This time interval was divided into regular subintervals of 16 s, which we call temporal bins. Good data statistics can be obtained by 
using this time bin width, because it represents approximately four full RHESSI rotations \citep{Lin2002}. In each temporal bin, the X-ray spectrum was 
reconstructed in units of photon flux using the Object SPectral EXecutive package ({\tt OSPEX}) \citep{Schwartz2002}. 

One crucial part of the impulsivity parameter calculation consists of separating both parts of the X-ray spectrum, the thermal and the 
nonthermal components \citep{Grigis2004}. In general, the thermal component is present at all flare stages and it is dominant only at low energies ($E<30$ keV), while the nonthermal
component has a transient behavior and appears mostly during the impulsive phase. 

So, in order to separate the two contributions, we need to choose a time interval during which we ensure that both are present. 
We solve this task taking a time interval of three minutes centered over the most prominent peak observed 
at the sum of the RHESSI channels 25-50 keV and 50-100 keV. The temporal location of this peak gives us an estimate 
of the time at which the most intense nonthermal emission takes place. This three-minute time interval spans 11 temporal bins.

\begin{figure}[htb]
\centering
\includegraphics[width=1.0\columnwidth]{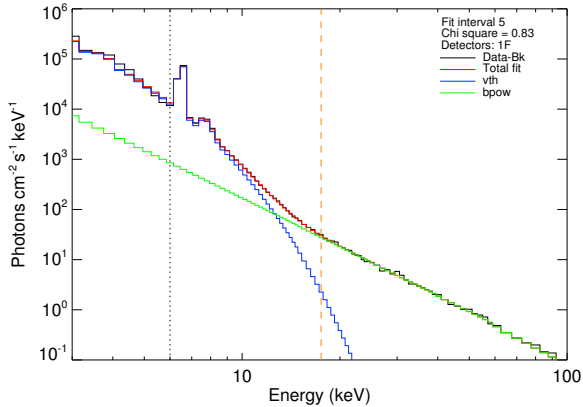}
\caption{{\scriptsize Reconstructed observational spectrum for the time interval from 17:53:28 to 17:53:44 UT of the flare (SOL2012-06-03T17:55-M3.3). 
Here are shown the two theoretical emission profiles, vth in blue and bpow in green. The whole theoretical fit appears in red and the observational
spectrum in black. The latter was fitted above 6 keV (vertical dotted black line) since the instrument attenuator state was A1. 
The orange dashed vertical line points out the transition energy found for this temporal bin, 17.5 $\pm$ 0.5 keV.}} 
\label{fig4.1} 
\end{figure}

The X-ray spectra corresponding to the 11 temporal bins mentioned previously are fitted using the same set of theoretical emission functions available in 
{\tt OSPEX}. This set is composed of a thermal function of one dominant temperature\footnote{The thermal function was used in full mode, i.e. it reproduces 
the thermal continuum plus the iron line centered at 6.7 keV.} (vth), and a broken power law with two different spectral indices (bpow). The sum of these 
gives the theoretical total profile, which is quite similar to the reconstructed spectrum made by RHESSI. We select the energy interval to fit 
in each temporal bin according to the respective attenuator state of the instrument. The intervals used were 3-100 keV, 6-100 keV, and 10-100 keV for 
attenuator states A0, A1, and A3, respectively. Figure \ref{fig4.1} shows the X-ray spectrum, its total fit, and the different theoretical functions 
for one temporal bin of our example flare. For our purposes, we always choose the same set of theoretical functions to analyze the whole flare sample. 
All of the function parameters were always allowed to vary.

Now, our interest is to estimate the energy that separates the thermal and nonthermal parts of the spectrum for each temporal bin. 
Hereafter, we will denote it as {\it transition energy}. 
We define this quantity as the energy at which the nonthermal component is 10 times higher (in photon flux per unit energy) than the thermal one. 
In other words, it marks the beginning of the nonthermal regime in the spectra.
As the X-ray emission from the flare evolves in time, the transition energy changes from one temporal bin to another. 
However, these variations are small and are confined in a narrow energy band.
Then, we take the transition energy for each flare as the average of the corresponding results for the 11 temporal bins.
In our example case (SOL2012-06-03T17:55-M3.3) the transition energy is 17.8 $\pm$ 0.4 keV. 

\begin{figure}[htb]
\centering
\includegraphics[width=1.0\columnwidth]{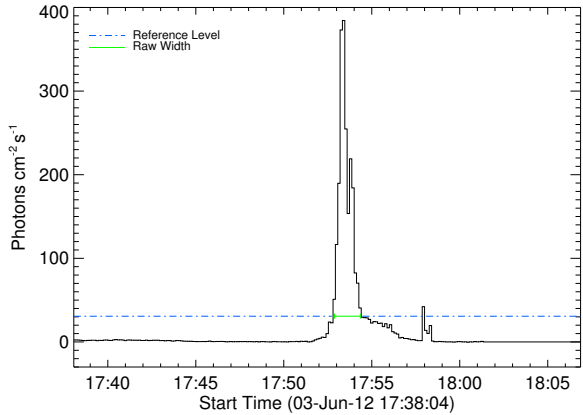}
\caption{{\scriptsize Lightcurve made from the nonthermal X-rays produced by the flare (SOL2012-06-03T17:55-M3.3). 
The horizontal axis spans the analysis time interval mentioned at the beginning of Section \ref{da}.
The blue dashed-dotted line indicates the 
reference flux level value. In this case, it was taken at 8\% of the maximum flux value. The green line segment points out the duration of the most 
prominent peak (104 s in this instance), which is inversely proportional to the impulsivity of this event.}} 
\label{fig4.2} 
\end{figure}

The next step is the creation of the lightcurve from the nonthermal X-rays. To do this, we take the reconstructed X-ray spectra for all of the initial temporal
bins and integrate them from the transition energy to 100 keV. 
We choose the transition energy as the lower integration limit because emissions coming from energies above this limit are dominated by the 
nonthermal component. Thus, we avoid thermal contributions.
Furthermore, we take 100 keV as the upper integration limit because most of the flares have extremely low photon 
fluxes at higher energies. The results of the integrations compose the lightcurve of interest in units of photon flux. 
Then, we located the most prominent peak in the lightcurve and measured its duration at the reference flux level (raw width) as is shown in Figure \ref{fig4.2}. This reference level 
is chosen ad hoc between 5\% and 10\% of the maximum lightcurve value. Thus, we avoid that prolonged emissions at low intensities can contribute 
to the measurement of the impulsive phase duration. Lastly, the ratio between the normalization factor ($t_0$) and the raw width ($t$) is the quantity 
that we consider as the impulsivity parameter of the flare (\textsc{IP})

\begin{equation}
\textsc{IP} = \frac{t_0}{t}
\label{ec2}
\end{equation}

The normalization factor can be defined as the mean of all the raw widths measured in the work sample. 
In our work sample, it has a value of 282.8 $\pm$ 0.7 s, but for practical purposes we select it ad hoc as 300 s.\footnote{All of the raw widths measured are 
integers, so we prefer to also choose an integer close to the average value of the impulsive phase duration as the normalization factor.}

In general, the most prominent nonthermal X-ray emission of a solar flare is characterized by having no symmetry. In fact, the impulsive phase duration 
can be separated in two parts, an increase and a decrease. The former would be the time between the beginning of the strong HXR emission until its peak, 
while the latter would be the time required for the flux to reach again the reference flux level. We want to know how similar these times are, thus we 
define the degree of symmetry ($S$) of a peak emission as

\begin{equation}
S = \frac{\textsc{DT}-\textsc{IT}}{\textsc{DT}+\textsc{IT}}
\label{ec3}
\end{equation}

Where, \textsc{DT} and \textsc{IT} are the decay and increase times, respectively. In summary, for our example event (SOL2012-06-03T17:55-M3.3) the raw width is 
104 $\pm$ 8 s, the degree of symmetry is 0.23 $\pm$ 0.06 and the impulsivity parameter is 2.885 $\pm$ 0.222. 
In Section 4, we will see how these data can be interpreted in comparison with the other sample results. 
The whole process described here was repeated for the other 47 events.

\section{Results}
\label{re}

\begin{figure}[htb]
\centering
\includegraphics[width=1.0\columnwidth]{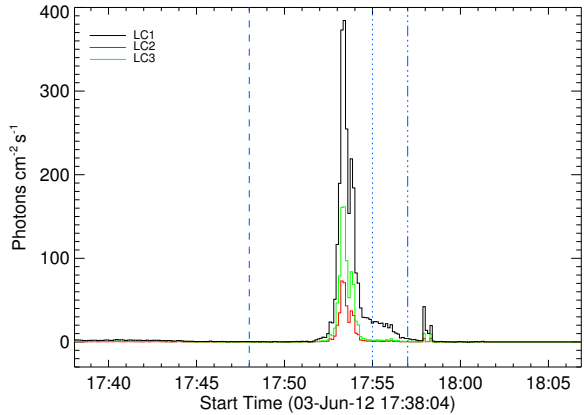}
\caption{{\scriptsize Lightcurves LC1, LC2, and LC3 for (SOL2012-06-03T17:55-M3.3) produced via the three methods of spectral integration mentioned earlier. 
The vertical dashed, dotted, and dashed-dotted lines represent the start, peak, and end times of the event according to GOES, respectively.
The impulsivity parameters for LC2 and LC3 have the same value, 3.125 $\pm$ 0.260. 
The reference flux level used to carry out the measurements was also 8\% of their corresponding maximum flux values.}}
\label{fig4.3} 
\end{figure}

So far, we have discussed one method to estimate the impulsivity of a solar flare; however, this is not the only way to do it. One can 
consider that the nonthermal X-ray lightcurve, from which the impulsivity parameter is calculated, can be made by integrating the X-ray spectra in fixed 
energy intervals (H. Hudson 2015, private communication). 
Carrying out the spectral integrations over a fixed energy interval will allow an easier reproduction of the impulsivity parameter results.
Following this idea, we create two other nonthermal X-ray lightcurves for each flare in the sample, using the energy intervals: 30-40 keV and 
25-50 keV. In these two new methods, we measured the impulsivity parameter in the same manner as before. Hereafter, the lightcurves made by 
integrating the X-ray spectra from the transition energy to 100 keV, from 30 to 40 keV and from 25 to 50 keV will be called LC1, LC2, and LC3, respectively.
Figure \ref{fig4.3} shows the three lightcurves generated for our example flare (SOL2012-06-03T17:55-M3.3). All of them are located between the 
start time and peak time of the flare according to GOES, as is expected from the Neupert effect.

In principle, the impulsivity can be calculated for all flares; therefore, we believe that it should work as a relevant feature over which these events can be
classified. Following this way of thinking, we propose a new system of classification for solar flares where we define three different types of 
impulsivity, namely, high, medium, and low. We arrange this system in such a manner that the events were distributed uniformly among the impulsivity types.
Table 1 shows the work sample classification for the three different results of the impulsivity parameter. 

\begin{table}
\centering
\begin{tabular}{lcrrr} \\\hline
Type & \textsc{IP} range & R1 & R2 & R3 \\\hline
High & \textsc{IP} $>2.0$ & $15$ & $17$ & $13$ \\
Medium & $1.0\leq$ \textsc{IP} $\leq2.0$ & $15$ & $16$ & $20$ \\ 
Low & \textsc{IP} $<1.0$ & $16$ & $12$ & $14$ \\
NC & --- & $2$ & $3$ & $1$ \\\hline
\end{tabular}                                      
\label{tb4.2.1} 
\caption{{\scriptsize Classification of the work sample according to the impulsivity parameter. R1, R2, and R3 are the results for lightcurves 
LC1, LC2, and LC3 respectively. NC: nonclassified events.}}
\end{table}


It was not possible to compute the impulsivity parameter for some flares in the work sample because their lightcurves did not show prominent peaks over the
background level or the noise was comparable with the signal. These flares only represent a few percent of the sample in any of the three 
analyzed methods. We refer to them as nonclassified events (NC).


Additionally, we divide each impulsivity type into three categories according to their degree of symmetry (S). 
We imposed the following conditions over the subclassification system: 
the thresholds between the categories of symmetry should be equidistant from $S=0$, and 
each impulsivity type must have at least one case of dominant injection and dominant decay.
Therefore, we define the categories of symmetry as 
dominant injection ($S <$ -0.2, INJ), symmetrical emission ($-0.2 \leq S \leq 0.2$, SYM) and dominant decay ($S >$ 0.2, DEC). 
This subdivision was done for the three impulsivity parameter results as appears in Table \ref{tb4.2.2}.

\begin{table}[ht]
\centering
\begin{tabular}{lrrrc} \hline
IP Type & R1 & R2 & R3 & S cat \\\hline
\multirow{3}{*}{High} & $1$ & $3$ & $0$ & INJ \\
                      & $9$ & $9$ & $9$ & SYM \\
                      & $5$ & $5$ & $4$ & DEC \\\hline
\multirow{3}{*}{Medium} & $3$ & $2$ & $2$ & INJ \\
                      & $6$ & $7$ & $7$ & SYM \\
                      & $6$ & $7$ & $11$ & DEC \\\hline
\multirow{3}{*}{Low}  & $2$ & $3$ & $0$ & INJ \\
                      & $2$ & $2$ & $5$ & SYM \\
                      & $12$ & $7$ & $9$ & DEC \\\hline
\end{tabular}                                      
\label{tb4.2.2} 
\caption{{\scriptsize Subdivision of the work sample classification taking into account the symmetry categories. 
R1, R2 and R3 as in Table 1.}}
\end{table}


We apply a Fisher exact test \citep{Fisher1922} to the contingency Table \ref{tb4.2.1}, without taking into account NC events, in order to confirm or to deny the following 
hypothesis: ``the alternative system of classification for solar flares is independent of the method chosen to measure the impulsivity parameter.''
The test gives a {\it p-value}  of 0.770, which supports the hypothesis. This result is above the significance level selected (0.10).\footnote{The null hypothesis can be denied whether the {\it p-value} is less than or equal to the significance level, otherwise it is confirmed.} 
This statistical tool was also applied to the three subsets of the work sample classification corresponding to each impulsivity type (3x3 blocks belonging to Table \ref{tb4.2.2}).
The {\it p-values} obtained were 0.699, 0.886, and 0.252 for high, medium, and low impulsivity types respectively. Therefore, we can conclude that the subdivision
of the classification system according to the degree of symmetry is independent with respect to the method used to measure the impulsivity parameter.


These correlations suggest to us that the three methods studied here to compute the impulsivity are congruent.
However, they present strong differences when the bulk nonthermal emission is centered at an energy that is not shared by the three different energy
integration intervals. This changes the morphology of the lightcurves considerably and therefore the calculations derived from them. 
This is the case of the flare (SOL2011-12-29T21:51-M2.0) whose impulsivity parameter results are 0.798 $\pm$ 0.017 for LC1, 2.500 $\pm$ 0.167 for LC2, 
and 1.014 $\pm$ 0.027 for LC3.

\begin{figure}[htb]
\centering
\includegraphics[width=1.0\columnwidth]{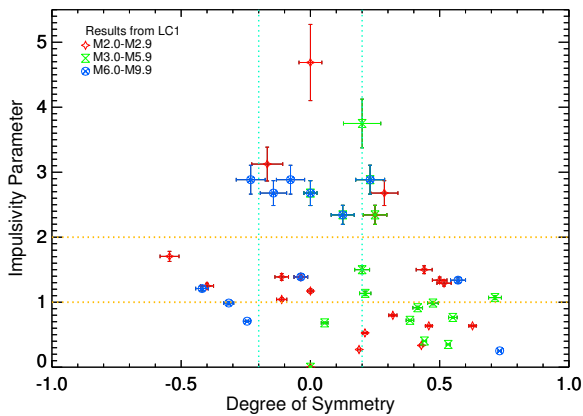}
\caption{{\scriptsize Distribution of the impulsivity parameter with respect to the degree of symmetry for R1 results. 
Nominal uncertainties of DT, IT, and the raw width were taken as 8s, i.e. the half duration of a temporal bin.
The corresponding error propagation treatment was calculated as is formulated in \citet{Ardila2007}. 
Horizontal and vertical dotted lines mark the thresholds for the impulsivity types and symmetry categories, respectively.}}
\label{fig4.4} 
\end{figure}

On the other hand, we wonder if the impulsivity parameter is a function of the degree of symmetry.
We approach this question by making scatter plots of such quantities for the three sets of obtained results.
In general, we find two relevant behaviors. First of all, we find a correlation between the impulsivity types and the categories 
of symmetry. Highly impulsive events have the most symmetrical emissions, and weakly impulsive ones are mainly characterized by having a dominant decay. 
The transition between these two sharp trends is given by medium impulsive flares, in which SYM and DEC categories are equally important.  
Therefore, the overall effect seen on the scatter plots (leaving out INJ events) is a broad band whose edges are located at the upper-center and 
bottom-right positions of the diagram. The majority of the events are located inside this band. This behavior can be seen in Figure \ref{fig4.4} and 
can also be inferred from Table \ref{tb4.2.2}. Second, these results suggest that events showing dominant injection are unlikely in this sample 
because prolonged injection times are rarely seen.

Additionally, the events were separated into smaller groups taking into account their GOES classes, in order to seek more restricted correlations.
Particularly, for results derived from LC1, flares ranging between GOES classes M3.0 and M5.9 are mainly located in the positive S-side of the scatter plot 
as is seen in Figure \ref{fig4.4}. 
We must bear in mind that these results could change if we include events of another GOES class (e.g. from M1.0 to M1.9). Hence, to consolidate 
the general relation between the impulsivity parameter and the degree of symmetry is required to apply the same analysis to a larger sample. 
We plan to do this in a later work.

\section{Discussion}
\label{fd}


A total of 48 flares were analyzed with a GOES classification ranging from M2.0 to M9.9, which occurred 
during the current Solar Cycle 24. The transition energy, lightcurves (LC1, LC2, and LC3) and impulsivity parameters were derived from the nonthermal X-ray 
emission data for each flare as is described in sections~\ref{da} and \ref{re}. The impulsivity parameter tell us how quickly the nonthermal X-ray 
emissions produced by a flare occur.


We propose that the duration of the impulsive phase of a solar flare can be considered proportional to the lifetime of 
the primary electron beam, under the assumptions that most of the HXR are released in a single pulse and footpoints are the main emitters. We think the former 
relation can be satisfied for highly impulsive events because they have the fastest and the most symmetrical emissions.
An alternative scenario is that there is ongoing acceleration throughout the flare \citep[e.g.][]{Miller1997,Fletcher2008}. 
This is supported by the existence of extended nonthermal emissions, i.e. medium and low impulsive events.
We believe both processes of HXR generation are present for flares with a dominant decay (DEC), where there is an initial stage with a fast release 
(primary beam) and then a continuous particle acceleration driven by another mechanism that produces HXR emissions of lower intensities.
It would be interesting to study in a deeper way which kind of physical processes could generate these emissions in all morphological regimes: symmetrical, 
dominant injection, and dominant decay. This is a very challenging task because the HXR lightcurves are taxonomically diverse.


The main advantage of the fixed energy integration interval methods (i.e. LC2 and LC3) is that they give fast results for the impulsivity parameter, and thus 
they are easy to reproduce. Nevertheless, their use is restricted to events where the nonthermal bulk emission must be within the studied energy interval. 
Also, if the local plasma is heated to high temperatures during the flare, the lightcurves LC2 and LC3 can be ``contaminated" by the thermal component of the 
spectrum, which is exactly what we would like to avoid. 
On the other hand, although the transition energy method is more complex than the others, it lets us cover the whole range of the nonthermal
X-ray emission regardless of its energy limits. However, this method depends strongly on the set of theoretical functions chosen to do the spectral
fits because the transition energy may vary by $\approx 5$ keV from one set of functions to another. It can be improved by repeating the fitting process 
with another set of theoretical functions, e.g. a multithermal function and the thin-kappa transport, and making the new transition energy as the mean value of 
the results obtained by using the two sets. In spite of the foregoing limitation, it is important to recognize the worth of the systematic way used here to separate 
both contributions of the X-ray spectrum produced by solar flares. A similar effort was carried out by \citet{Grigis2004}.


One of the main results of the present work is the alternative system of classification for solar flares. It is based on a feature that is common to all flares and
is directly related to their temporal evolution. This system, as well as its subdivision, according to the degree of symmetry, seems to be independent with respect to 
the method chosen to compute the impulsivity parameter. Unfortunately, we are limited to analyze events that have been observed by RHESSI, which 
significantly reduces the number of events that can be studied. 


The sample used to build this classification system does not take into account all GOES flare types; hence it must be applied with caution to 
flares outside of the considered class range (M2.0 to M9.9). Because the system is still under development, it could be improved by including a significant number 
of events outside of the current sample. In such cases, we would count with more statistics, and the classification would be more general and reliable.


Another important result is the correlation found between the impulsivity parameter and the degree of symmetry. 
This was described as a broad band seen in the scatter plots made from both quantities.
According to the previous discussion about HXR generation, this correlation suggests us which kind of particle acceleration mechanism (primary electron beam or acceleration 
throughout the flare) is relevant for each impulsivity type. 
We want to check whether this correlation holds for a bigger sample. 
If so, we plan to describe the main trend in greater detail, by introducing the function IP$=$IP($S$). 
A similar procedure can be performed for the data dispersion from this trend, $\sigma=\sigma$(IP,$S$).

 
We consider both the total amount of energy input and the impulsivity as relevant features that can define later behaviors in the evolution of solar 
flares. The former is believed to be emitted mostly in white light and during the impulsive phase \citep{Kretzschmar2011}. The ratio between both would be an 
estimate of the energy input rate. If it is high enough, we believe that other kinds of impulsive solar activity can manifest in the solar medium, such as 
the EIT-waves, Moreton waves, or sunquakes. These phenomena can only be achieved by very impulsive (IP$>2.0$) and highly energetic flares.
 

In the future, we plan to consolidate our work in three different ways. First, we will increase the number of events in the work 
sample by taking into account the flares that occurred in Solar Cycle 23. Thereby, we will count with a greater sample; 
thus, our system of classification of solar flares could be improved and become more representative. 
Second, we plan to correlate the impulsivity parameter with other observational features 
such as the Neupert effect and temporal evolution of the spectral index. We want to know in what proportion this effect and this evolutionary pattern holds
in each type of impulsivity. Finally, we will explore the use of the energy input rate as a diagnostic tool of other kinds of impulsive 
solar phenomena. We hope to apply the methods explained here and check the results already shown by using data from future {solar X-ray observations}.

The authors wish to acknowledge the important contribution from Hugh Hudson, who proposed the alternative methods (LC2 and LC3) for the impulsivity parameter calculation. Also, 
we thank to Lindsay Glesener for her wonderful and patient guidance regarding the use of the RHESSI software and the transition energy estimations. 
Finally, we are deeply grateful to Charles Lindsey and to the anonymous referee, whose kind and timely comments considerably improved the manuscript. 
J.C.M.-O. was supported by NASA under contract NNX11AP05G.

\bibliographystyle{apj}
\bibliography{IPSF}

\begin{thebibliography}{34}
\expandafter\ifx\csname natexlab\endcsname\relax\def\natexlab#1{#1}\fi

\bibitem[{{Abbett} \& {Hawley}(1999)}]{Abbett1999}
{Abbett}, W.~P. \& {Hawley}, S.~L. 1999, The Astrophysical Journal, 521, 906

\bibitem[{{Antonucci} {et~al.}(1982){Antonucci}, {Gabriel}, {Acton},
  {Leibacher}, {Culhane}, {Rapley}, {Doyle}, {Machado}, \&
  {Orwig}}]{Antonucci1982}
{Antonucci}, E., {Gabriel}, A.~H., {Acton}, L.~W., {Leibacher}, J.~W.,
  {Culhane}, J.~L., {Rapley}, C.~G., {Doyle}, J.~G., {Machado}, M.~E., \&
  {Orwig}, L.~E. 1982, Solar Physics, 78, 107

\bibitem[{{Ardila}(2007)}]{Ardila2007}
{Ardila}, A.~M. 2007, F\'isica Experimental (2nd ed.; Unibiblos, Universidad
  Nacional de Colombia)

\bibitem[{{Benz}(2002)}]{Benz2002}
{Benz}, A., ed. 2002, Astrophysics and Space Science Library, Vol. 279, {Plasma
  Astrophysics. Kinetic Processes in Solar and Stellar Coronae}, ed. A.~{Benz}
  (Kluwer Academic Publishers), 7--11

\bibitem[{Benz(2008)}]{Benz2008}
Benz, A. 2008, Living Reviews in Solar Physics, 5, 5

\bibitem[{{Brown}(1971)}]{Brown1971}
{Brown}, J.~C. 1971, Solar Physics, 18, 489

\bibitem[{{Cargill} \& {Priest}(1983)}]{CargillPriest1983}
{Cargill}, P.~J. \& {Priest}, E.~R. 1983, Astrophysical Journal, 266, 383

\bibitem[{{Crannell} {et~al.}(1978){Crannell}, {Frost}, {Saba}, {Maetzler}, \&
  {Ohki}}]{Crannell1978}
{Crannell}, C.~J., {Frost}, K.~J., {Saba}, J.~L., {Maetzler}, C., \& {Ohki}, K.
  1978, The Astrophysical Journal, 223, 620

\bibitem[{{Dennis} \& {Schwartz}(1989)}]{Dennis1989}
{Dennis}, B.~R. \& {Schwartz}, R.~A. 1989, Solar Physics, 121, 75

\bibitem[{{Dennis} \& {Zarro}(1993)}]{Dennis1993}
{Dennis}, B.~R. \& {Zarro}, D.~M. 1993, Solar Physics, 146, 177

\bibitem[{{Emslie}(1978)}]{Emslie1978}
{Emslie}, A.~G. 1978, Astrophysical Journal, 224, 241

\bibitem[{{Emslie} {et~al.}(2012){Emslie}, {Dennis}, {Shih}, {Chamberlin},
  {Mewaldt}, {Moore}, {Share}, {Vourlidas}, \& {Welsch}}]{Emslie2012}
{Emslie}, A.~G., {Dennis}, B.~R., {Shih}, A.~Y., {Chamberlin}, P.~C.,
  {Mewaldt}, R.~A., {Moore}, C.~S., {Share}, G.~H., {Vourlidas}, A., \&
  {Welsch}, B.~T. 2012, The Astrophysical Journal, 759, 71

\bibitem[{{Fisher}(1922)}]{Fisher1922}
{Fisher}, R.~A. 1922, Journal of the Royal Statistical Society, 85, 87

\bibitem[{{Fletcher} {et~al.}(2013){Fletcher}, {Hannah}, {Hudson}, \&
  {Innes}}]{Fletcher2013}
{Fletcher}, L., {Hannah}, I.~G., {Hudson}, H.~S., \& {Innes}, D.~E. 2013, The
  Astrophysical Journal, 771, 104

\bibitem[{{Fletcher} \& {Hudson}(2008)}]{Fletcher2008}
{Fletcher}, L. \& {Hudson}, H.~S. 2008, The Astrophysical Journal, 675, 1645

\bibitem[{{Glesener} {et~al.}(2013){Glesener}, {Krucker}, {Bain}, \&
  {Lin}}]{Glesener2013}
{Glesener}, L., {Krucker}, S., {Bain}, H.~M., \& {Lin}, R.~P. 2013, \apjl, 779,
  L29

\bibitem[{{Graham} {et~al.}(2011){Graham}, {Fletcher}, \&
  {Hannah}}]{Graham2011}
{Graham}, D.~R., {Fletcher}, L., \& {Hannah}, I.~G. 2011, Astronomy \&
  Astrophysics, 532, A27

\bibitem[{{Graham} {et~al.}(2013){Graham}, {Hannah}, {Fletcher}, \&
  {Milligan}}]{Graham2013}
{Graham}, D.~R., {Hannah}, I.~G., {Fletcher}, L., \& {Milligan}, R.~O. 2013,
  The Astrophysical Journal, 767, 83

\bibitem[{{Grigis} \& {Benz}(2004)}]{Grigis2004}
{Grigis}, P.~C. \& {Benz}, A.~O. 2004, Astronomy and Astrophysics, 426, 1093

\bibitem[{{Hirayama}(1974)}]{Hirayama1974}
{Hirayama}, T. 1974, Solar Physics, 34, 323

\bibitem[{Hudson(2011)}]{Hudson2011}
Hudson, H. 2011, Space Science Reviews, 158, 5

\bibitem[{{Hudson}(1991)}]{Hudson1991}
{Hudson}, H.~S. 1991, in Bulletin of the American Astronomical Society,
  Vol.~23, Bulletin of the American Astronomical Society, 1064

\bibitem[{{Kiplinger}(1995)}]{Kiplinger1995}
{Kiplinger}, A.~L. 1995, Astrophysical Journal, 453, 973

\bibitem[{{Kretzschmar}(2011)}]{Kretzschmar2011}
{Kretzschmar}, M. 2011, Astronomy \& Astrophysics, 530, A84

\bibitem[{{Krucker} \& {Lin}(2008)}]{Krucker2008}
{Krucker}, S. \& {Lin}, R.~P. 2008, The Astrophysical Journal, 673, 1181

\bibitem[{{Lin} {et~al.}(2002){Lin}, {Dennis}, {Hurford}, {Smith}, {Zehnder},
  {Harvey}, {Curtis}, {Pankow}, {Turin}, {Bester}, {Csillaghy}, {Lewis},
  {Madden}, {van Beek}, {Appleby}, {Raudorf}, {McTiernan}, {Ramaty}, {Schmahl},
  {Schwartz}, {Krucker}, {Abiad}, {Quinn}, {Berg}, {Hashii}, {Sterling},
  {Jackson}, {Pratt}, {Campbell}, {Malone}, {Landis}, {Barrington-Leigh},
  {Slassi-Sennou}, {Cork}, {Clark}, {Amato}, {Orwig}, {Boyle}, {Banks},
  {Shirey}, {Tolbert}, {Zarro}, {Snow}, {Thomsen}, {Henneck}, {McHedlishvili},
  {Ming}, {Fivian}, {Jordan}, {Wanner}, {Crubb}, {Preble}, {Matranga}, {Benz},
  {Hudson}, {Canfield}, {Holman}, {Crannell}, {Kosugi}, {Emslie}, {Vilmer},
  {Brown}, {Johns-Krull}, {Aschwanden}, {Metcalf}, \& {Conway}}]{Lin2002}
{Lin}, R.~P., {Dennis}, B.~R., {Hurford}, G.~J., {Smith}, D.~M., {Zehnder}, A.,
  {Harvey}, P.~R., {Curtis}, D.~W., {Pankow}, D., {Turin}, P., {Bester}, M.,
  {Csillaghy}, A., {Lewis}, M., {Madden}, N., {van Beek}, H.~F., {Appleby}, M.,
  {Raudorf}, T., {McTiernan}, J., {Ramaty}, R., {Schmahl}, E., {Schwartz}, R.,
  {Krucker}, S., {Abiad}, R., {Quinn}, T., {Berg}, P., {Hashii}, M.,
  {Sterling}, R., {Jackson}, R., {Pratt}, R., {Campbell}, R.~D., {Malone}, D.,
  {Landis}, D., {Barrington-Leigh}, C.~P., {Slassi-Sennou}, S., {Cork}, C.,
  {Clark}, D., {Amato}, D., {Orwig}, L., {Boyle}, R., {Banks}, I.~S., {Shirey},
  K., {Tolbert}, A.~K., {Zarro}, D., {Snow}, F., {Thomsen}, K., {Henneck}, R.,
  {McHedlishvili}, A., {Ming}, P., {Fivian}, M., {Jordan}, J., {Wanner}, R.,
  {Crubb}, J., {Preble}, J., {Matranga}, M., {Benz}, A., {Hudson}, H.,
  {Canfield}, R.~C., {Holman}, G.~D., {Crannell}, C., {Kosugi}, T., {Emslie},
  A.~G., {Vilmer}, N., {Brown}, J.~C., {Johns-Krull}, C., {Aschwanden}, M.,
  {Metcalf}, T., \& {Conway}, A. 2002, Solar Physics, 210, 3

\bibitem[{{Masuda} {et~al.}(1994){Masuda}, {Kosugi}, {Hara}, {Tsuneta}, \&
  {Ogawara}}]{Masuda1994}
{Masuda}, S., {Kosugi}, T., {Hara}, H., {Tsuneta}, S., \& {Ogawara}, Y. 1994,
  Nature, 371, 495

\bibitem[{{Miller} {et~al.}(1997){Miller}, {Cargill}, {Emslie}, {Holman},
  {Dennis}, {LaRosa}, {Winglee}, {Benka}, \& {Tsuneta}}]{Miller1997}
{Miller}, J.~A., {Cargill}, P.~J., {Emslie}, A.~G., {Holman}, G.~D., {Dennis},
  B.~R., {LaRosa}, T.~N., {Winglee}, R.~M., {Benka}, S.~G., \& {Tsuneta}, S.
  1997, Journal of Geophysical Research,, 102, 14631

\bibitem[{{Neupert}(1968)}]{Neupert1968}
{Neupert}, W. 1968, The Astrophysical Journal, 153, L59

\bibitem[{{Ning}(2007)}]{Ning2007}
{Ning}, Z. 2007, The Astrophysical Journal, 659, L69

\bibitem[{{Schwartz} {et~al.}(2002){Schwartz}, {Csillaghy}, {Tolbert},
  {Hurford}, {McTiernan}, \& {Zarro}}]{Schwartz2002}
{Schwartz}, R.~A., {Csillaghy}, A., {Tolbert}, A.~K., {Hurford}, G.~J.,
  {McTiernan}, J., \& {Zarro}, D. 2002, Solar Physics, 210, 165

\bibitem[{{Shibata}(1996)}]{Shibata1996}
{Shibata}, K. 1996, Advances in Space Research, 17, 9

\bibitem[{{Veronig} {et~al.}(2005){Veronig}, {Brown}, {Dennis}, {Schwartz},
  {Sui}, \& {Tolbert}}]{Veronig2005}
{Veronig}, A.~M., {Brown}, J.~C., {Dennis}, B.~R., {Schwartz}, R.~A., {Sui},
  L., \& {Tolbert}, A.~K. 2005, The Astrophysical Journal, 621, 482

\bibitem[{{Zhang} {et~al.}(2001){Zhang}, {Dere}, {Howard}, {Kundu}, \&
  {White}}]{Zhang2001}
{Zhang}, J., {Dere}, K.~P., {Howard}, R.~A., {Kundu}, M.~R., \& {White}, S.~M.
  2001, The Astrophysical Journal, 559, 452

\end{thebibliography}

\end{document}